\documentclass[12pt]{article}
\usepackage{amsmath}
\usepackage{amsfonts}
\begin{document}

\vspace*{1cm}

\begin{center}
{\bf{\Large A Proposal for the Vector State in Vacuum \\
\vspace*{.35cm}

String Field Theory }}

\vspace*{1cm}

Radoslav Rashkov\footnote{e-mails: rrachkov@sfu.ca; rash@phys.uni-sofia.bg,
on leave of absence from Dept. of Physics, Sofia University, 1164 Sofia,
Bulgaria}
and K Sankaran Viswanathan\footnote{e-mail: kviswana@sfu.ca}\\

\ \\
Department of Physics, Simon Fraser University \\
Burnaby, BC, V5A 1S6, Canada
\end{center}

\vspace*{.8cm}

\begin{abstract}
A previous calculation on the tachyon state arising as fluctuations
of a $D$ brane in vacuum string field theory is extended to include
the vector state. We use the boundary conformal field theory approach
of Rastelli, Sen and Zwiebach to construct a vector state. It is shown
that the vector field satisfies the linearized equations of motion provided 
the two conditions $k^2=0$ and $k^\mu A_\mu=0$ are satisfied. Earlier
calculations using Fock space techniques by Hata and Kawano have found
massless vector states that are not necessarily transverse.
\end{abstract}

\vspace*{.8cm}

\section{Introduction}

After the significant progress in understanding of the tachyon condensation
in open string theory, Rastelli, Sen and Zwiebach proposed the so called
Vacuum String Field Theory (VSFT) as a candidate for the theory of open string
dynamics around the tachyon vacuum \cite{rsz1}. Regarded as a significant step
toward solving Witten's open string field theory \cite{w}, for the last one 
year it has been intensively studied \cite{rsz2,gt1,gt2,mt,rsz5,rsz3}. 
Being constructed solely of world-sheet ghost fields, the kinetic operator 
$\mathcal{Q}$
in VSFT  is non-dynamical \cite{rsz}. The exact relation between Open
String Field Theory (OSFT) and VSFT is not completely clear but there are 
significant developments in this direction \cite{mt,grsz,ki-oh,ok}. 
An important step toward a deeper understanding of VSFT is the study of the
fluctuation modes around  a classical solution to the VSFT called 
the sliver, and
the proof that the known physical spectrum of $D$-branes can be reproduced.
The work in this direction was initiated by Hata and Kawano who proposed a
Fock space construction of the tachyon and vector states around the
tachyon vacuum \cite{hk,hm}. In a very interesting paper \cite{rsz4} 
Rastelli, Sen and Zwiebach re-investigated the Hata-Kawano tachyon 
state using boundary conformal field theory (BCFT) and pointed out to 
the difficulties in the computation of the $D$25-brane tension.  In 
our previuos paper \cite{rv} we reinvestigated the tachyon state in 
boundary conformal field theory language and showed that the BPZ inner 
product between slivers must be treated carefully.  We proposed a 
consistent prescription for calculating both the star product of two 
slivers and their BPZ product and showed that the tachyon state 
satisfies the linearized equations of motion in the strong sense as 
well as in weak sense.

In \cite{hk} a proposal for vector state has been also given in framework of
Fock space formalism. The authors of \cite{hk} found the correct on-shell
conditions for the vector field on the brane, but they fail to find the
transversality condition and the polarization vector is arbitrary
rather than satisfying the transversality condition. It was pointed out in 
\cite{ki-oh} that within this framework it would be a hard task
to impose transversality condition on the vector or tensor massless fields. 
In this short note we propose a vector state in analogy to the tachyon state 
by making use of BCFT approach. To find the on-shell conditions for the vector 
field $|V\rangle$ we follow the procedure used in \cite{rsz4} and \cite{rv}
for the tachyon field $|\chi_T\rangle$. By making use of BCFT prescription
we are able to find the correct on-shell conditions for the vector field and
more importantly, its transversality.

\section{BCFT construction of the vector state}

The open string field theory (OSFT) action proposed by Witten is given by
\cite{w}
\begin{equation}
S=-\kappa\left\{\frac 12\langle\Psi|\,Q\,|\Psi\rangle+\frac 13
\langle\Psi|\Psi\star\Psi\rangle\right\},
\label{1}
\end{equation}
where $|\Psi\rangle$ is the string field represented by a ghost number one
state in the matter-ghost BCFT, $\langle\cdot |\cdot \rangle$ represents the
BPZ inner product and $\star$ is the string star multiplication. In
(\ref{1}) $Q$ is the ordinary BRST charge made up from 
energy-momentum tensors for matter and ghost sectors, $T_{matter}$ and
$T_{ghost}$, and the ghost as well. The equations of motion following from
(\ref{1}) are
\begin{equation}
Q|\Psi\rangle+|\Psi\star\Psi\rangle=0.
\label{2}
\end{equation}
After the open string tachyon condensation, the theory must represent closed 
string vacuum and there should be no open string perturbative excitations.
VSFT is proposed to describe the dynamics around the tachyon vacuum and
its kinetic operator $\mathcal{Q}$ must have trivial cohomology. 
The action for VSFT is given by
\begin{equation}
S=-\kappa\left\{\frac 12\langle\Psi|
\mathcal{Q}|\Psi\rangle+\frac 13
\langle\Psi|\Psi\star\Psi\rangle\right\},
\label{3}
\end{equation}
where $\mathcal{Q}$ is a new BRST operator of ghost number one and 
made of ghost fields:
\begin{equation}
\mathcal{Q}=c_0+\sum\limits_{n\geq 1}f_n[c_n+(-1)^nc^\dagger_n]
\label{4}
\end{equation}
where $f_n$ are some numerical coefficients. Since $\mathcal{Q}$ has this 
special form, the solutions are factorized as follows
\begin{equation}
|\Psi\rangle=|\Psi_g\rangle\otimes|\Psi_m\rangle ,
\label{5}
\end{equation}
where $|\Psi_m\rangle$  and $|\Psi_g\rangle$ are the
matter and ghost part respectively. The equations of motion then 
take the following form
\begin{align}
& \mathcal{Q}|\Psi_g\rangle=|\Psi_g\star\Psi_g\rangle \notag \\
& |\Psi_m\rangle=|\Psi_m\star\Psi_m\rangle.
\label{6}
\end{align}
The ghost part is taken to be universal, but $|\Psi_m\rangle$ is a solution
of projector equation and represents a particular brane.
This interpretation follows from
the fact that with $\mathcal{Q}$ constructed purely of ghost fields, 
it has trivial
cohomology and hence solutions to VSFT contain no perturbative open string
states, but may describe non-perturbative states such as the D-branes. 
The properties and the solutions to VSFT have been discussed by
Rastelli, Sen and Zwiebach in several papers \cite{rsz1,rsz2,rsz3,rsz}.
The approach using the so called surface states turn out to be 
very useful since one can represent the BPZ inner product and 
the star product by simple geometric
operations and using BCFT. The solution to the matter part of the equations of
motion describing a $D$25-brane, called a sliver, is defined through the
relation \cite{rsz3,rsz}
\begin{equation}
\langle\Xi_m|\Phi\rangle= \lim_{n\to\infty}
\mathcal{N}
\langle f\circ\Phi(0)\rangle_{C_n},
\label{7}
\end{equation}
where $C_n$ is a semi infinite cylinder of circumference $\frac{n\pi}{2}$ 
obtained by making the identification
 ${\Re}\,z\simeq{\Re}\,z+\frac{n\pi}{2}$ in the upper half $z$ plane.
The map $f$ is given by $f(z)=tan^{-1}z$, $|\Phi\rangle$ is an arbitrary state 
in the matter Hilbert space and
$\langle\dots\rangle_{C_n}$ denotes the correlation function of the matter
BCFT on the semi infinite cylinder $C_n$. In the limit $n\to\infty$, $C_n$ 
approaches the upper half $z$ plane. The sliver state is
normalized so that
\begin{equation}
\langle\Xi_m|\Xi_m\rangle=KV^{(26)}, \qquad (V^{(26)}=(2\pi)^{26}
\delta^{26}(0))
\label{8}
\end{equation}
where $V^{(26)}$ is the volume of 26-dimensional spacetime and $K$ is a
normalization constant that arises due to anomaly in the matter sector.

The tachyon state can be written in the form \cite{hk,rsz4,rv}
\begin{equation}
|\Psi_g\rangle\otimes|\chi_T(k)\rangle
\label{9}
\end{equation}
where $|\Psi_g\rangle$ is the same state for all brane solutions and
$|\chi_T(k)\rangle$ is the matter part defined as
\begin{equation}
\langle\chi_T(k)|\Psi\rangle=\mathcal{N}\lim_{n\to\infty} n^{2k^2}
\langle e^{ik.X(\frac{n\pi}{4})}f\circ\Psi(0)\rangle_{C_n} 
\label{10}
\end{equation}
for any state  $|\Psi\rangle$ in the Hilbert space of states of the string 
field\footnote{For more detais about the general construction see
\cite{rsz3,rsz}.}. It has been shown that the state $|\chi_T\rangle$ 
satisfies the linearized equations of motion in weak sense \cite{rsz4} and
in strong sense as well \cite{rv}.

We propose the following form of the vector state on $D$25-brane
\begin{equation}
\langle V(k)|\Psi\rangle=\mathcal{N}\lim_{n\to\infty} n^{2(k^2+1)}
\langle A_\mu\partial_t X^\mu(\frac{n\pi}{4}) e^{ik.X(\frac{n\pi}{4})}
f\circ\Psi(0)\rangle_{C_n} 
\label{11}
\end{equation}
i.e. we insert a vector vertex operator in the middle of the wedge state,
diametrically opposite to the puncture at the origin, and in the limit
$n\to\infty$ the wedge state approaches the sliver. The subscript in 
$\partial_t$ is for tangential derivative to the boundary. 
The normalization factors are chosen in such a way that the BPZ product 
with Fock states is nonvanishing.
As we mentioned above, BPZ inner product and $\star$-product of two slivers
can be constructed by cutting and gluing cylinders $C_n$ and $C_{n'}$ as
prescribed in \cite{rsz3,rsz}. The expression for $f\circ\Psi$ can be
expanded as 
\begin{equation}
f\circ\Psi(0)=a_\Psi \left[ e^{-ik.X(0)}\right]+
\left[\zeta_\nu \partial_t X^\nu(0) e^{-ik.X(0)}\right] + \cdots
\label{12}
\end{equation}
where square brackets denote the conformal blocks of primary
operators. The different conformal blocks in (\ref{12}) are orthogonal
to each other. With these definitions we can calculate
(\ref{11}) following the procedures outlined in \cite{rsz4,rv}.

\section{The linearized equations and transversality}

In this section we will show that the state defined in (\ref{11}) 
has the correct on-shell condition for massless vector field and, in addition,
it is transverse. Using the equation (\ref{12}) we can write (\ref{11}) as
\begin{align}
\langle V(k)|\Psi\rangle=&\lim_{n\to\infty}\mathcal{N}a_\Psi n^{2(k^2+1)}
\left\{
A_\mu(k)\langle\partial X^\mu e^{ik.X(\frac{n\pi}{4})}
\,e^{-ik.X(0)} \rangle_{C_n} \right. \notag \\
&+\left. \zeta_\nu(-k)A_\mu(k)
\langle\partial X^\mu e^{ik.X(\frac{n\pi}{4})}
\,\partial X^\nu e^{-ik.X(0)} \rangle_{C_n} \right\}
\label{13}
\end{align}
where we take into account only the primary operators for the first two
conformal blocks, terms containing $\partial X^\mu$ hereafter should be 
understood as $\partial_t X^\mu$ and the argument will always be that 
of the exponent following $\partial X$.  As argued in \cite{rsz3}, the 
descendents and the higher conformal blocks are more suppressed in 
$n\to\infty$ limit and therefore their contributions vanish.

The second term in the curly brackets in (\ref{13}) is
\begin{multline}
\lim_{n\to\infty}\mathcal{N}a_\Psi n^{2(k^2+1)}
\zeta_\nu(-k)A_\mu(k)
\langle\partial X^\mu e^{ik.X(\frac{n\pi}{4})}
\partial X^\nu(0) e^{-ik.X(0)} \rangle_{C_n}
\\
=\lim_{n\to\infty}\left[\mathcal{N}a_\Psi n^{2(k^2+1)}
\left(\frac 4n\right)^{2(k^2+1)}
\zeta_\nu(-k)A_\mu(k)\right. 
\\
\left. \langle\partial X^\mu(-1) e^{ik.X(-1)}
\partial X^\nu(0) e^{-ik.X(0)} \rangle_D\right](2\pi)^{26}\delta(0)
\\ 
=-\mathcal{N}a_\Psi A_\mu\zeta_\nu \eta^{\mu\nu}
 2^{2k^2+3}\, V^{(26)}
\\
\label{14}
\end{multline}
where the second equality is found after mapping $C_n$ to unit disk and 
computing the correlations functions on the disk from the BCFT.

The first term in (\ref{11}) can be evaluated in the same way and we find
\begin{multline}
\lim_{n\to\infty}\mathcal{N}a_\Psi n^{2(k^2+1)}
A_\mu(k)\langle\partial X^\mu e^{ik.X(\frac{n\pi}{4})}
 e^{-ik.X(0)} \rangle_{C_n}
\\
=\lim_{n\to\infty}\left[\mathcal{N}a_\Psi n^{2(k^2+1)}
\left(\frac 4n\right)^{(2k^2+1)}iA_\mu(k) 
\langle\partial X^\mu(-1) e^{ik.X(-1)}
 e^{-ik.X(0)} \rangle_D\right]
\\ 
=\lim_{n\to\infty}\mathcal{N}a_\Psi n A_\mu k^\mu cotan(\frac{\pi}{2})
 2^{(2k^2+1)}(2\pi)^{26}\delta(0)= 0.
\\
\label{15}
\end{multline}
and therefore there is no contribution from this conformal block.

We are now in a position to prove that the vector state $|V\rangle$ (\ref{11})
satisfies the linearized equations of motion. To do so we have to calculate 
the quantity $\langle V\star\Xi+\Xi\star V|\Psi\rangle$ for arbitrary
$|\Psi\rangle$. According to \cite{rsz3,rsz4,rv} we have

\begin{align}
\langle \Xi\star V+ V\star\Xi|\Psi\rangle=&
\lim_{n\to\infty}\mathcal{N}a_\Psi n^{2(k^2+1)}
\bigl\{\bigr.
A_\mu\bigl[\bigr.
\langle\partial X^\mu e^{ik.X(\frac{n\pi}{4})}\,e^{-ik.X(0)}
\rangle_{C_{2n-1}}
\notag \\
&
\langle\partial X^\mu e^{ik.X(\frac{(3n-2)\pi}{4})}\,e^{-ik.X(0)}
\rangle_{C_{2n-1}}
\bigl. \bigr] \notag \\
&
+A_\mu\zeta_\nu\bigl[\bigr.
\langle\partial X^\mu e^{ik.X(\frac{n\pi}{4})}
\,\partial X^\nu e^{-ik.X(0)}
\rangle_{C_{2n-1}} 
\notag \\
& +
\langle\partial X^\mu e^{ik.X(\frac{(3n-2)\pi}{4})}
\,\partial X^\nu e^{-ik.X(0)}
\rangle_{C_{2n-1}}\bigl. \bigr]
\bigl.\bigr\}
\label{16}
\end{align}
where, according to the rules outlined in \cite{rv}, in the star product we 
take the semi infinite cylinders representing $|V\rangle$ and $|\Xi\rangle$
to be of equal circumference $n\pi/2$. After mapping to unit disk $D$ we find
\begin{align}
\langle \Xi\star V+ V\star\Xi|\Psi\rangle=&
\lim_{n\to\infty}\mathcal{N}a_\Psi n^{2(k^2+1)}
\bigl\{
A_\mu\bigl[
\langle\partial X^\mu e^{ik.X(i)}e^{-ik.X(0)}
\rangle_D
\bigr. \bigr.\notag \\
&
+\bigl.
\langle\partial X^\mu e^{ik.X(-i)}e^{-ik.X(0)}
\rangle_D
\bigr] \notag \\
&
+A_\mu\zeta_\nu
\bigl[
\langle\partial X^\mu e^{ik.X(i)}
\,\partial X^\nu e^{-ik.X(0)}
\rangle_D \bigr.
\notag \\
& +\bigl.\bigl.
\langle\partial X^\mu e^{ik.X(-i)}
\,\partial X^\nu e^{-ik.X(0)}
\rangle_D\bigr]
\bigr\}\, V^{(26)}
\notag \\
=&
-2\mathcal{N}a_\Psi \left[\eta^{\mu\nu}+k^\mu k^\nu\right]
\,\zeta_\mu A_\mu 2^{k^2+2}\, V^{(26)}
\notag \\
=& -\mathcal{N}a_\Psi \left[\eta^{\mu\nu}+k^\mu k^\nu\right]
\,\zeta_\mu A_\mu 2^{k^2+3}\, V^{(26)}.
\label{17}
\end{align}
where the correlation functions in the first square brackets cancel
each other and the evaluation of the correlators in the second
brackets gives the final result.
But from (\ref{14}) and (\ref{15}) we have
\begin{equation}
\langle V|\Psi\rangle=
-\mathcal{N}a_\Psi \zeta_\nu A_\mu
\eta^{\mu\nu} 2^{2k^2+3}V^{(26)}.
\label{18}
\end{equation}
The expression (\ref{18}) is equal to (\ref{17}) if
\begin{equation}
k^2=0 
\label{19}
\end{equation}
and
\begin{equation}
k^\mu A_\mu=0
\label{20}
\end{equation}
for any Fock state $|\Psi\rangle$ and thus the equations of motion
give the correct on-shell conditions
for the massless vector field $A_\mu$. We point out that along with $k^2=0$ we
found that the vector field $A_\mu$ is transverse which the authors of 
\cite{hk} failed to find using oscilator approach.

To obtain the kinetic term for $A_\mu$ in the action we must first check that
the vector state $|V\rangle$ satisfies the linearized equations of
motion against another sliver (which was a problem in \cite{rsz3} but a
resolution was proposed in \cite{rv}). According to the general
procedure developed in \cite{rsz3}, we have
\begin{align}
\langle V\star\Xi|\chi\rangle=&
\lim_{n_i\to\infty}n_1^{2{k'}^2}n_2^{2(k^2+1)}A_\mu
\times \notag \\
&\langle\partial X^\mu e^{ik.X[(2n_2+n_3+n_1-4)\pi/4]}e^{ik'.X(0)}
\rangle_{C_{n_1+n_2+n_3-3}}.
\label{21}
\end{align}
To compute (\ref{21}) we follow the procedure outlined in \cite{rv} and
first map the first strip to a strip of length $\pi/2$. The length of the 
other two strips become $\tilde n_2=\frac{n_2}{n_1-1}$ and
$\tilde n_3=\frac{n_3}{n_1-1}$, and we define also $\epsilon_1=
\frac{1}{n_1-1}$. Since we have the star product $\langle V\star\Xi|$, we 
take $\tilde n_2=\tilde n_3=\tilde n$. A simple computation yields
\begin{align}
\langle V\star\Xi|\chi\rangle=&
\lim_{n_1,\tilde n\to\infty}\tilde n i A_\mu 2^{{k'}^2+k^2+1}
\langle\partial X^\mu e^{ik.X(3\pi/2)}e^{ik'.X(0)}
\rangle_D 
\notag \\
=&-\lim_{\tilde n\to\infty} \tilde n ik^\mu A_\mu 2^{k^2+1}V^{(26)}
\label{22}
\end{align}
and taking into account that the result for $\langle\Xi\star V|\chi\rangle$
is the same but with opposite sign, we find
\begin{equation}
\langle\Xi\star V+ V\star\Xi |\chi\rangle=0.
\label{23}
\end{equation}
Analogous calculations yield
\begin{equation}
\langle V\star\Xi+\Xi\star V|V\rangle=
-(\eta^{\mu\nu}+k^\mu k^\nu)A_\mu A_\nu 2^{k^2+3}V^{(26)}
\label{24}
\end{equation}
and
\begin{align}
&\langle V|\chi\rangle=\lim n k^\mu A_\mu 2^{2(k^2+1)}
cotan(\frac{\pi}{2})V^{(26)}=0,
\label{25} \\
&\langle V|V\rangle=-\eta^{\mu\nu} A_\mu A_\nu 2^{2k^2+3}
V^{(26)}.
\label{26}
\end{align}
Comparing (\ref{23}-\ref{26}) we conclude that the linearized equations
of motion are satisfied also against slivers under conditions
(\ref{19}) and (\ref{20}).

To compute the kinetic term in the action (\ref{3}), we define the off-shell
tachyon field $T(k)$ and vector field $A_\mu(k)$ in the momentum space
in the following way
\begin{multline}
|\Psi\rangle=|\Psi_g\rangle\otimes\Bigl[
|\Xi_m\rangle \\
+\int d^{26}k\left(
n^{-k^2}T(k)|\chi_T(k)\rangle +n^{-k^2-1}A_\mu(k)|V^\mu(k)\rangle 
+\dots \right)\Bigr]
\label{27}
\end{multline}
where ellipses denote of higher order excitations. Substitution of
(\ref{27}) into (\ref{3}) gives for the quadratic part in $A_\mu$ 
\begin{align}
S^{(2)}=&-\frac 12\langle\Psi_g|\mathcal{Q}|\Psi_g\rangle
\int d^{26}k_1d^{26}k_2\, A_\mu(k_1)A_\nu(k_2)\, n_1^{-k_1^2-1}
n_2^{-k_1^2-1}
\notag \\
&\langle V^\mu(k_1)|\left[V^\nu(k_2)\rangle -
|\Xi_m\star V^\nu(k_2)\rangle -|V^\nu(k_2)\star\Xi\rangle
\right].
\label{28}
\end{align}
The BPZ products in the above expression are readily computed in (\ref{21}-
\ref{26}) and we have simply to substitute their values into (\ref{28}).
Then for the kinetic term we get
\begin{align}
S^{(2)}&=-\frac 12\langle\Psi_g|
\mathcal{Q}|\Psi_g\rangle \int d^{26}k_1\,d^{26}k_1\,
n_1^{-k_1^2-1}\,n_2^{-k_2^2-1} A_\mu(k_1)\, A_\nu(k_2)
\notag \\
&\left[
\langle V^\mu(k_1)|V^\nu(k_2)\rangle -
\langle V^\mu(k_1)|\Xi_m\star V^\nu(k_2)\rangle -
\langle V^\mu(k_1)|V^\nu(k_2)\star\Xi\rangle
\right]
\notag \\
& \simeq
-\frac 12\langle\Psi_g|
\mathcal{Q}|\Psi_g\rangle\,\int d^{26}k_1\,d^{26}k_2\, 
n_1^{-k_1^2-1}\,n_2^{-k_2^2-1}\, k_1^2 \, A_\mu(k_1)\, A_\nu(k_2) \notag 
\\
&
\qquad \, \left\{ \eta^{\mu\nu}\, 2^{2k_1^2+3} -(\eta^{\mu\nu} -k_1^\mu 
k_2^\mu)2^{k_1^2+3}\right\}\delta(k_1+k_2) \notag \\
& \simeq
-\frac{\langle\Psi_g|
\mathcal{Q}|\Psi_g\rangle}{2}2^3\,ln2\,\int d^{26}k
(k^2\eta^{\mu\nu}-\frac{k^\mu k^\nu}{ln\,2}) 
A_\mu(k)\, A_\nu(-k)
\label{30}
\end{align}
where in the second equality we take the vector field near on-shell and
in the third step we used the procedure outlined in \cite{rv} to
calculate the correlation function keeping terms quadratic in $k_\mu$.
It is clear from (\ref{30}) that the expected 
action for the vector field on $D$25-brane reproduced from VSFT.
From (\ref{30}) we conclude that the proposed vector state gives the correct
on-shell conditions and, though $A_\mu$ is transverse, the correct action for 
a massless vector field.

Due to the twist symmetry of the VSFT action, the interaction part will contain
only terms linear in $\chi_T$ and quadratic in $V$ and they can be 
easily calculated using the above technique.
For this terms we find in coordinate space
\begin{equation}
S^{(2)}\sim 3\int d^{26}x T A_\mu A^\mu +
2\int d^{26}x \left[\partial_\mu\partial_\nu T A^\mu A^\nu 
-\partial_\mu T\partial_\nu A^\mu A^\nu +
T\partial_\nu  A^\mu \partial_\mu A^\nu\right]
\label{31}
\end{equation}
where, as in the case of tachyon interaction term, the multiplicative
constant is infinite.

\section{Conclusions}

In this note we have proposed a vector state around the classical solution of
VSFT representing a $D$25-brane. A careful analysis shows that the 
proposed state
satisfies the linearized equations of motion with the correct on-shell
conditions for a massless vector field $A_\mu$. We also find the transversality
condition on $A_\mu$, a property that seems difficult to 
demonstrate in the oscilator approach (see \cite{hk,hm}). 
Due to the twist symmetry, the interaction term must contain two vector field 
and one tachyon and we found it to be of the form (\ref{31}). 
Terms with two tachyons and one vector field do not appear. This is because 
the twist operation $\Omega$, i.e a combined world-sheet parity reversal 
plus $SL(2,R)$ transformations, is a symmetry of the OSFT and VSFT as well
\footnote{Seee for review for instance \cite{oh,ir}.}. Under the action of
$\Omega$ the massless vector field $A_\mu$ is odd and changes sign while
the tachyon field, being even, remains unchanged. For instance, one can 
easily compute $T(k_1)\,T(k_1)\,A_\mu(k_3)$ term:
\begin{align}
\langle\chi|\chi\star V\rangle
=&\lim_{n_i\to\infty}N\,n_1^{2k_1^2}\,n_2^{2k_2^2}\,n_3^{2(k_3^2+1)}
A_\mu(k_3) \notag \\
&\langle e^{ik_1.X(0)}e^{ik_2.X[(n_1+n_2-2)\pi/4]}
\partial X^\mu e^{ik_3.X[(n_1+2n_2+n_3-4)\pi/4]}
\rangle_{C_{n_1+n_2+n_3-3}}.
\label{32}
\end{align}
To compute (\ref{32}), we perform the following steps

a) map the first strip to $[-\pi/4,\pi/4]$.

b) change the variable $w=e^{i4\pi/(2n+\epsilon_1-3)}$ to map to unit disk
\footnote{Here $n=\frac{n_2}{n_1-1}=\frac{n_3}{n_1-1}$ and $\epsilon_1=
\frac{1}{n_1-1}$.}.

c) take $n$ and $n_1$ large and compute the correlator.

We end up with:
\begin{align}
\langle\chi|\chi\star V\rangle&=
\lim_{n,n_1\to\infty}N\,n^{k_2^2+k_3^2-k_1^2}\, 2^{k_1^2+k_2^2+k_3^2}A_\mu(k_3)
\notag \\
& 
\langle e^{ik_1.X(e^0)}\,e^{ik_2.X(e^{2i\pi/2})}
\,\partial X^\mu e^{ik_3.X(e^{3i\pi/2})}\rangle_D
\notag \\
&
\sim i\,(k_2-k_1)^\mu A_\mu(k_3)2^{k_1^2-1} 
(2\pi)^{26}\delta(k_1+k_2+k_3)
\label{33}
\end{align}
which is what is expected for this coupling.\footnote{See for instance 
\cite{gsw}.}
It would be interseting to investigate the fluctuation modes around the other 
$D$-brane solutions of VSFT and coincident $D$-branes as well. We hope to 
report on these issues in the near future.

\vspace*{.8cm}

{\bf Acknowledgements:}  R.R. would like to thank Simon Fraser
University for warm hospitality. This work has been supported by an
operating grant from the Natural Sciences and Engineering Research Council
of Canada.

\end{document}